\documentclass[conference]{IEEEtran}

\usepackage{cite}
\usepackage[mode=tex]{standalone}
\usepackage{amsmath,amssymb}
\usepackage{balance}
\usepackage{booktabs}
\usepackage{graphicx}
\usepackage{microtype}
\usepackage{pgfplots}
\pgfplotsset{compat=1.18}
\usepackage{textcomp}
\usepackage{xcolor}
\usepackage{xurl}
\usepackage[hidelinks]{hyperref}
\hypersetup{hidelinks,pdfborder={0 0 0}}
\usetikzlibrary{shapes.geometric}

\usepackage[nolist]{acronym}

\begin{document}
\newacro{pr}[PR]{protection relay}
\newacro{ml}[ML]{machine learning}
\newacro{rl}[RL]{reinforcement learning}
\newacro{cql}[CQL]{conservative Q-learning}
\newacro{mv}[MV]{medium voltage}
\newacro{pmu}[PMU]{phasor measurement unit}
\newacro{dft}[DFT]{discrete Fourier transform}
\newacro{td}[TD]{temporal difference}
\newacro{hif}[HIF]{high-impedance fault}
\newacro{tp}[TP]{true positive}
\newacro{tn}[TN]{true negative}
\newacro{fp}[FP]{false positive}
\newacro{fn}[FN]{false negative}
\newacro{fpr}[FPR]{false-positive rate}
\newacro{nft}[NFT]{non-fault false-trip rate}
\newacro{ci}[CI]{confidence interval}
\newacro{cpu}[CPU]{central processing unit}
\newacro{cuda}[CUDA]{Compute Unified Device Architecture}

\title{Offline Reinforcement Learning for Distribution-Grid Protection}

\author{
\IEEEauthorblockN{
Julian Oelhaf\textsuperscript{1\textdagger}\textsuperscript{*},
Alexander Luce\textsuperscript{1\textdagger},
Christian Bergler\textsuperscript{2},
Andreas Maier\textsuperscript{1},
Siming Bayer\textsuperscript{1}
}
\IEEEauthorblockA{
\textit{\textsuperscript{1}Pattern Recognition Lab, Friedrich-Alexander-Universit\"at Erlangen--N\"urnberg, Erlangen, Germany} \\
\textit{\textsuperscript{2}Department of Electrical Engineering, Media and Computer Science,} \\
\textit{Ostbayerische Technische Hochschule Amberg-Weiden, Amberg, Germany} \\
\textsuperscript{*}Corresponding author: julian.oelhaf@fau.de
}
}

\maketitle

\setcounter{footnote}{0}
\renewcommand{\thefootnote}{\fnsymbol{footnote}}
\footnotetext{\textdagger\ These authors contributed equally to this work.}
\renewcommand{\thefootnote}{\arabic{footnote}}

\begin{abstract}
Data-driven protection may complement conventional relays in distribution grids whose operating conditions vary with distributed generation, switching events, and changing short-circuit levels. We study line-selective tripping from static trajectories of a realistically simulated CIGRE medium-voltage network using offline reinforcement learning. A convolutional Q-network receives causal voltage-current phasor and apparent-impedance features, optionally together with raw waveforms, and is trained with conservative Q-learning (CQL). A controlled sensitivity study evaluates two observation windows, reward variants, and three CQL weights under a common split and training protocol; one exploratory post-hoc run additionally increases the discount factor from $\gamma$ = 0.95 to 0.99. On 225 held-out episodes, the best per-timestep result is obtained with combined input and CQL weight $\alpha$ = 0.9, reaching precision 0.9993, recall 0.9496, and F1-score 0.9738. Because dense per-timestep scores do not encode the terminal semantics of relay operation, we also evaluate the first non-wait action in each episode. The default combined-input agent selects the correct line-trip action first in 98.13\,\% of 214 fault episodes, but trips in 72.73\,\% of the 11 non-fault episodes. In the post-hoc run, the corresponding rates are 98.60\,\% and 54.55\,\%, respectively. The results show that dense predictive performance and terminal protection behavior can lead to different model rankings. Offline CQL therefore demonstrates strong faulted-line selection on the simulated fault episodes, while the static trajectories, small non-fault set, and single-seed post-hoc design preclude conclusions about practical relay security or deployment readiness.
\end{abstract}

\begin{IEEEkeywords}
conservative Q-learning, distribution-grid protection, offline reinforcement learning, protective relaying
\end{IEEEkeywords}

\section{Introduction}

Protection devices must isolate faults rapidly and selectively while remaining secure during switching and other benign transients. Conventional overcurrent schemes can be challenged by bidirectional power flows, changing short-circuit levels, and operating states introduced by distributed energy resources. These developments have motivated both data-driven fault analysis~\cite{yoon_development_2024} and \ac{rl}-based protection schemes, ranging from relay coordination~\cite{kilickiran_reinforcement_2018} to direct breaker control in closed-loop grid simulations~\cite{kordowich_hybrid_2022}.

The data are precomputed EvEMTBench transient simulations~\cite{kordowich_simulation_2026}, rather than interactions with a live grid. The task is therefore offline \ac{rl}: the agent must estimate relay-action values from a fixed dataset and cannot improve unsafe decisions through further exploration. \Ac{cql} addresses this setting by suppressing unsupported action values~\cite{kumar_conservative_2020}; related work has applied physics-guided \ac{cql} to generator tripping~\cite{gao_conservative_2025}.

\subsection{Related Work and Positioning}

Data-driven protection has predominantly been studied through supervised
fault detection, classification, and localization~\cite{oelhaf_scoping_2025}.
These approaches map voltage or current measurements to fault labels or
protection decisions using raw waveforms or phasor-domain quantities. Yoon
and Yoon employ segmented voltage waveforms in a Transformer-based model
for fault and disturbance classification~\cite{yoon_development_2024},
while Shanmugapriya and Baskaran use synchronized phasor measurements from
\acp{pmu} with a deep residual network~\cite{shanmugapriya_rapid_2023}.
Related controlled work has evaluated machine-learning methods for fault
detection and line identification under a common experimental
protocol~\cite{oelhaf_systematic_2025}. Such objectives can yield strong
predictive performance, but they generally treat observations as classification
samples and do not directly encode the asymmetric consequences of waiting,
tripping a healthy line, or selecting the wrong line.

\ac{rl} has been investigated for protection coordination and breaker control. Wu et al. formulate protective relaying as an interactive multi-agent \ac{rl} problem~\cite{wu_nested_2019}. Kordowich et al. similarly employ closed-loop PowerFactory simulations, but combine centralized teacher agents with decentralized backup agents~\cite{kordowich_hybrid_2022}. In both cases, breaker actions alter the subsequent trajectory, unlike the fixed trajectories considered here.

Offline \ac{rl} addresses learning from such fixed datasets but is vulnerable to overestimating weakly represented actions. \ac{cql} mitigates this effect through conservative value estimation~\cite{kumar_conservative_2020}. Gao et al. apply this principle to physics-guided generator-tripping control~\cite{gao_conservative_2025}, whereas the present work considers line-selective distribution-grid protection from static fault and switching-event trajectories.

This leaves a methodological gap between supervised protection models, which are typically evaluated as repeated classification decisions, and interactive \ac{rl} formulations, in which actions affect future states. Our setting lies between these cases: the policy is learned from a fixed archive, but its operational output is a terminal line-trip decision. The evaluation of such offline protection policies when dense per-timestep predictions and the first irreversible trip lead to different conclusions has received limited attention.

We address that gap with a controlled study of offline \ac{cql} on static CIGRE medium-voltage trajectories. We compare causal phasor and impedance features with combined raw-waveform inputs, and vary observation length, reward design, and \ac{cql} regularization under a common split and training protocol. Beyond per-timestep precision, recall, F1-score, and false-positive rate, we evaluate the first non-wait action in each episode to distinguish correct operation, wrong-line trips, missed trips, nuisance trips, and operating delay.

\section{System, Data, and Method}
\label{sec:system_method}

\subsection{Simulated Distribution-Grid Episodes}

\begin{figure}[ht]
\centering
\resizebox{0.96\columnwidth}{!}{%
  \begingroup

\tikzset{
  wire/.style={draw=black,line width=0.55pt,line cap=round,line join=round},
  bus/.style={draw=black,line width=2.15pt,line cap=rect},
  device/.style={draw=black,line width=0.75pt,line cap=round,line join=round},
  breaker/.pic={
    \fill (-0.10cm,-0.10cm) rectangle (0.10cm,0.10cm);
  },
  open breaker/.pic={
    \draw[device,line width=1.05pt]
      (-0.08cm,-0.08cm) rectangle (0.08cm,0.08cm);
  },
  load/.pic={
    \draw[wire] (0cm,0cm) -- (0cm,-0.18cm);
    \pic at (0cm,-0.28cm) {breaker};
    \draw[wire] (0cm,-0.38cm) -- (0cm,-0.62cm);
    \draw[device] (-0.17cm,-0.62cm) -- (0.17cm,-0.62cm)
      -- (0cm,-0.91cm) -- cycle;
  },
  transformer/.pic={
    \draw[device] (0cm,0.12cm) circle (0.16cm);
    \draw[device] (0cm,-0.12cm) circle (0.16cm);
  },
  converter/.pic={
    \draw[wire] (0cm,0cm) -- (0cm,0.13cm);
    \pic at (0cm,0.23cm) {breaker};
    \draw[wire] (0cm,0.33cm) -- (0cm,0.47cm);
    \pic at (0cm,0.67cm) {transformer};
    \draw[wire] (0cm,0.95cm) -- (0cm,1.08cm);
    \draw[device] (-0.18cm,1.08cm) rectangle (0.18cm,1.44cm);
    \draw[device] (-0.13cm,1.35cm) -- (-0.01cm,1.24cm)
      -- (-0.10cm,1.24cm) -- (0.11cm,1.13cm);
    \draw[device] (-0.13cm,1.18cm) -- (-0.03cm,1.29cm)
      -- (0.13cm,1.18cm);
  }
}

\begin{tikzpicture}[x=0.72cm,y=1cm]
  \path[use as bounding box] (-0.15,-0.25) rectangle (18.15,13.75);

  \begin{scope}[shift={(8.5,13.03)}]
    \begin{scope}[x=1cm,y=1cm]
      \begin{scope}
        \clip (-0.38,-0.38) rectangle (0.38,0.38);
        \foreach \d in {-1.2,-0.8,...,1.2} {
          \draw[device] (-0.8,\d-0.8) -- (0.8,\d+0.8);
          \draw[device] (-0.8,\d+0.8) -- (0.8,\d-0.8);
        }
      \end{scope}
      \draw[device,line width=1.15pt] (-0.38,-0.38) rectangle (0.38,0.38);
    \end{scope}
  \end{scope}
  \draw[wire] (8.5,12.65) -- (8.5,12.34);
  \pic at (8.5,12.10) {transformer};
  \draw[wire] (8.5,11.82) -- (8.5,11.00);

  \draw[bus] (3.00,11.00) -- (12.00,11.00);

  \draw[wire] (4.00,11.00) -- (4.00,10.42);
  \pic at (4.00,10.16) {transformer};
  \draw[wire] (4.00,9.88) -- (4.00,9.47);
  \pic at (4.00,9.37) {breaker};
  \draw[wire] (4.00,9.27) -- (4.00,9.15);
  \draw[bus] (3.00,9.15) -- (5.00,9.15);
  \pic at (3.28,9.15) {load};
  \pic at (4.00,8.96) {breaker};

  \draw[wire] (4.00,9.15) -- (4.00,7.80);
  \pic at (4.00,7.90) {breaker};
  \draw[bus] (3.08,7.68) -- (5.08,7.68);
  \pic at (4.00,7.46) {breaker};
  \draw[wire] (4.00,7.68) -- (4.00,6.47);
  \pic at (4.00,6.56) {breaker};
  \draw[bus] (1.20,6.32) -- (9.25,6.32);
  \pic at (5.75,6.32) {load};
  \pic at (8.10,6.32) {converter};

  \draw[wire] (1.90,6.32) -- (1.90,5.47);
  \pic at (1.90,6.13) {breaker};
  \pic at (1.90,5.57) {breaker};
  \draw[bus] (1.20,5.34) -- (9.25,5.34);
  \pic at (8.55,5.34) {load};

  \draw[wire] (1.90,5.34) -- (1.90,3.20);
  \pic at (1.90,5.13) {breaker};
  \pic at (1.90,3.30) {breaker};
  \draw[bus] (1.40,3.06) -- (3.38,3.06);
  \pic at (1.90,2.86) {breaker};
  \pic at (3.05,3.06) {load};

  \draw[wire] (7.20,5.34) -- (7.20,4.18);
  \pic at (7.20,5.13) {breaker};
  \pic at (7.20,4.28) {breaker};
  \draw[bus] (6.60,4.08) -- (9.00,4.08);
  \pic at (7.20,3.88) {breaker};
  \pic at (8.65,4.08) {load};
  \draw[wire] (7.20,4.08) -- (7.20,3.08);
  \pic at (7.20,3.20) {breaker};
  \draw[bus] (6.58,2.94) -- (9.00,2.94);
  \pic at (7.20,2.73) {breaker};

  \draw[wire] (11.30,11.00) -- (11.30,10.42);
  \pic at (11.30,10.16) {transformer};
  \draw[wire] (11.30,9.88) -- (11.30,9.54);
  \pic at (11.30,9.43) {breaker};
  \draw[wire] (11.30,9.33) -- (11.30,9.23);
  \draw[bus] (10.72,9.23) -- (13.72,9.23);
  \pic at (13.38,9.23) {load};

  \draw[wire] (11.30,9.23) -- (11.30,7.60);
  \pic at (11.30,8.95) {breaker};
  \pic at (11.30,7.70) {breaker};
  \draw[bus] (10.70,7.47) -- (13.75,7.47);
  \pic at (11.30,7.25) {breaker};
  \pic at (13.40,7.47) {load};

  \draw[wire] (11.30,7.47) -- (11.30,5.82);
  \pic at (11.30,5.92) {breaker};
  \draw[bus] (10.62,5.68) -- (18.00,5.68);
  \pic at (11.30,5.46) {breaker};
  \pic at (14.75,5.68) {converter};
  \pic at (17.50,5.68) {load};

  \draw[wire] (11.30,5.68) -- (11.30,3.28);
  \pic at (11.30,3.57) {open breaker};
  \pic at (11.30,3.12) {breaker};
  \draw[bus] (10.62,2.98) -- (15.02,2.98);
  \pic at (14.72,2.98) {load};

  \draw[wire] (13.55,2.98) -- (13.55,1.97);
  \pic at (13.55,2.76) {breaker};
  \pic at (13.55,2.08) {breaker};
  \draw[bus] (12.72,1.86) -- (15.10,1.86);
  \pic at (13.55,1.64) {breaker};
  \pic at (14.72,1.86) {load};

  \draw[wire] (13.55,1.86) -- (13.55,0.86);
  \pic at (13.55,0.98) {breaker};
  \draw[bus] (12.72,0.74) -- (15.10,0.74);
  \pic at (13.55,0.52) {breaker};
  \pic at (14.72,0.74) {load};

  \draw[wire] (7.20,2.94) -- (7.20,2.02) -- (9.55,2.02);
  \pic at (9.38,2.02) {breaker};
  \draw[wire] (9.55,2.02) -- (11.30,2.02) -- (11.30,2.98);
  \pic at (9.78,2.02) {breaker};
  \draw[wire,line width=1.5pt] (9.57,1.54) -- (9.57,2.55);
  \draw[wire] (9.57,1.28) -- (9.57,1.66);
  \pic at (9.75,1.32) {breaker};
  \draw[wire] (9.85,1.32) -- (10.07,1.32);
  \draw[device] (10.07,1.16) -- (10.07,1.48) -- (10.34,1.32) -- cycle;

  \draw[wire] (1.90,3.06) -- (1.90,0.15) -- (13.55,0.15) -- (13.55,0.74);
\end{tikzpicture}
\endgroup%
}
\caption{CIGRE benchmark medium-voltage grid in the European configuration.}
\label{fig:cigre-grid}
\end{figure}
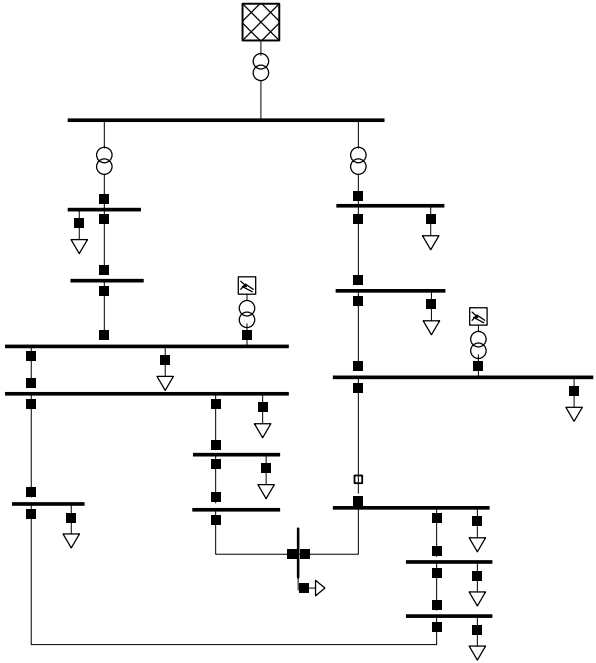

The CIGRE medium-voltage benchmark network in its European configuration~\cite{cigre_task_force_c60402_benchmark_2014}, shown in Fig.~\ref{fig:cigre-grid}, is simulated with DIgSILENT PowerFactory~\cite{digsilent_gmbh_powerfactory_2024}. Trajectories are taken from the EvEMTBench dataset~\cite{kordowich_simulation_2026}. The network has 14 buses and 15 line segments. Measurements are recorded at both terminals of each line except one cubicle at MainBus~8 on line 8--14, yielding 29 line cubicles. Three-phase currents and voltages therefore provide 174 raw channels per timestep. We use 4{,}507 episodes of $0.5$~s sampled at $9.6$~kHz. This subset includes nine fault families, including single-phase-to-ground (1ph-G), 2ph, 2ph-G, and 3ph faults, and eleven non-fault switching or disturbance types. Processed episodes align events at $0.1$~s. The agent receives neither a time index nor an onset indicator; alignment is used only for labels and latency evaluation.

The held-out evaluation set contains 225 episodes (214 fault and 11 non-fault); the other 4,282 episodes form the development pool. The transition archive contains 1,686,758 rows sampled before onset, densely around onset, and sparsely thereafter. Excluding incomplete phasor windows, 1,656,784 rows remain for window size $W=48$ and 1,626,810 for $W=96$. The development episodes are divided with seed zero into 3,853 optimization episodes and 429 internal monitoring episodes; no episode crosses a partition boundary. A comparison with the current EvEMTBench metadata showed that the project dataset contains 4{,}507 simulations with IDs 0--4506, whereas the cleaned online metadata contains 4{,}509 simulations. The two additional records, IDs 4507 and 4508, are non-fault \texttt{switch\_ibr\_trip} events and were not included in training or evaluation. For the 4{,}507 shared simulations, event types and targets match exactly. The experiments therefore use 4{,}353 fault and 154 non-fault episodes, whereas the complete cleaned metadata contains 156 non-fault episodes. Differences in event-time origin, arc-time-constant units, and directory prefixes are consistent with differing metadata and file-location conventions; they do not change the event-class assignments used here.

\begin{table}[t]
\centering
\caption{Summary of the dataset subset used for training and evaluation.}
\label{tab:data}

\setlength{\tabcolsep}{4pt}
\begin{tabular}{lr}
\toprule
\textbf{Property} & \textbf{Value} \\
\midrule
Network buses / line segments & 14 / 15\\
Line cubicles / raw channels & 29 / 174\\
Episodes: development / held-out & 4,282 / 225\\
Held-out: fault / non-fault & 214 / 11\\
Episode duration / sampling rate & 0.5 s / 9.6 kHz\\
Event onset / grid frequency & 0.1 s / 50 Hz\\
Trip actions / wait actions & 15 / 1 \\
\bottomrule
\end{tabular}
\end{table}

\subsection{Causal Signal Representation}

Raw inputs contain the 174 line-cubicle waveform channels. Under normal operation, these voltage and current waveforms oscillate at the 50-Hz fundamental. Phasors provide a more interpretable fundamental-frequency representation, while apparent impedance may supply a physically motivated inductive bias for line-selective protection. The phasor representation applies a full-cycle sliding discrete Fourier transform using $N=9600/50=192$ timesteps. For signal $p$,

\begin{equation*}
P[n]=\frac{2}{N}\sum_{k=0}^{N-1}p[n-k]e^{-j2\pi k/N}.
\label{eq:phasor}
\end{equation*}

Voltage and current phasors give the apparent impedance

\begin{equation*}
Z_{c,\phi}[n]=\frac{U_{c,\phi}[n]}{I_{c,\phi}[n]}
             =R_{c,\phi}[n]+jX_{c,\phi}[n],
\label{eq:impedance}
\end{equation*}

for cubicle $c$ and phase $\phi$ \cite{horowitz_power_2023}. Division by very low current magnitudes is suppressed using the causal reference

\begin{equation*}
m_c[n]=\max_{\tau\leq n}\max_{\phi\in\{a,b,c\}}
       |I_{c,\phi}[\tau]|.
\label{eq:causal-peak}
\end{equation*}

All experiments use this causal maximum; a full-trajectory maximum would leak future information. The impedance is retained when $|I_{c,\phi}[n]|\geq0.005m_c[n]$ and replaced by zero otherwise. For each phase, the model receives $|U|$, $|I|$, $R$, and $X$, totaling 348 phasor features. The combined representation uses separate convolutional branches for the 174 raw and 348 phasor channels.

At decision timestep $n$, the state $s_n$ is a matrix containing the most recent $W$ timesteps across the $C$ input channels, i.e., $s_n\in\mathbb{R}^{W\times C}$. Thus, features at $n$ use only measurements through $n$. This finite window approximates the Markov property because longer dependencies cannot be excluded. We evaluate observation lengths $W\in\{48,96\}$ feature timesteps, corresponding to one-quarter and one-half of a 50-Hz cycle at the feature-sequence level. Because each phasor uses a trailing 192-sample cycle, the causal raw-signal support spans 239 samples for $W=48$ and 287 samples for $W=96$. The action space has 16 choices: trip both breakers associated with one of the 15 line segments, or wait.

\subsection{Offline Conservative Q-Learning}

The sequential protection task is formulated as a finite-horizon Markov decision process $\mathcal{M}=(\mathcal{S},\mathcal{A},P,R,\gamma)$~\cite{sutton_reinforcement_2018}. Here, $\mathcal{S}$ contains the finite-window measurement states defined above, $\mathcal{A}$ comprises the 15 line-trip actions and the wait action, $R$ assigns the protection-dependent rewards in \eqref{eq:reward}, and $\gamma$ discounts future rewards. In an interactive protection environment, the transition model $P(s'|s,a)$ would describe how a trip or wait action changes the subsequent electrical state. The available archive is static, however: $s'$ is the next recorded window in the PowerFactory trajectory and is not regenerated in response to the selected action. The problem is therefore an offline, action-independent approximation of the interactive decision process~\cite{levine_offline_2020}.

A dilated one-dimensional convolutional network maps each observation to 16 Q-values. Each input has shape $W\times C$, is batch-normalized channel-wise, and is rearranged to the $(B,C,W)$ layout required by the one-dimensional convolutions. For combined input, the raw-waveform and phasor branches are pooled separately, concatenated, and passed to a fully connected action head.

With the default settings, the reward is

\begin{equation}
r(s,a)=
\begin{cases}
5, & \text{correct fault trip after onset},\\
5, & \text{wait after onset in a non-fault episode},\\
0, & \text{wait before onset or during a fault},\\
-100, & \text{pre-event, non-fault, or wrong-line trip}.
\end{cases}
\label{eq:reward}
\end{equation}

The heuristic reward values reflect the asymmetric cost of disconnecting a healthy line or selecting the wrong line. Zero reward for waiting before onset or during a fault allows the policy to defer its decision without an immediate penalty, while the positive reward reinforces secure post-event waiting and correct selective tripping. Counterfactual wrong-line actions augment the fixed transition set.

The temporal-difference loss uses the target network $Q_{\bar{\theta}}$:

\begin{equation*}
\mathcal{L}_{\mathrm{TD}}=
\mathbb{E}_{(s,a,r,s')\sim\mathcal{D}}
\left[
\left(
Q_\theta(s,a)
-r
-\gamma\max_{a'}Q_{\bar{\theta}}(s',a')
\right)^2
\right].
\label{eq:td-loss}
\end{equation*}

\Ac{cql} adds the conservative regularization term

\begin{equation*}
\mathcal{L}_{\mathrm{CQL}}=
\mathbb{E}_{s\sim\mathcal{D}}
\left[
\log\sum_a e^{Q_\theta(s,a)}
-Q_\theta(s,a_{\mathcal{D}})
\right],
\label{eq:cql-loss}
\end{equation*}

where $a_{\mathcal{D}}$ denotes the action contained in the offline dataset. The total objective is
\begin{equation*}
\mathcal{L}
=
\mathcal{L}_{\mathrm{TD}}
+
\alpha\mathcal{L}_{\mathrm{CQL}}.
\label{eq:total-loss}
\end{equation*}

The counterfactual examples provide direct supervision for selected unsafe actions, but they cover only a subset of the possible state--action combinations. The CQL term complements this augmentation by suppressing high Q-values for actions that are weakly supported by the offline data distribution. At inference, the policy selects

\begin{equation*}
\pi_\theta(s)=\arg\max_{a\in\mathcal{A}}Q_\theta(s,a).
\label{eq:policy}
\end{equation*}

All prespecified models are trained for 30 epochs with Adam at $10^{-3}$, batch size 1024, $\gamma=0.95$, and $\tau=0.005$ on one A100 MIG GPU with 16 CPU cores; the 853{,}604-parameter combined model trains in about 63~min. The post-hoc model differs only in $\gamma=0.99$. The 429-episode internal monitoring partition is used to track training losses but not to select checkpoints or compute the reported performance metrics; those metrics use the separate 225-episode evaluation set. Random-number generators and data loading use seed zero; strict deterministic CUDA kernels are not enforced.

\begin{table}[ht]
\centering
\caption{Model and training settings.}
\label{tab:training}
\begin{tabular}{lr}
\toprule
\textbf{Setting} & \textbf{Value}\\
\midrule
Phasor channels & 348\\
Combined channels & 174 raw + 348 phasor\\
Convolution kernel / dilations & 7 / 1, 3, 9, 27\\
Pooled features per branch & 128\\
Optimizer / learning rate & Adam / $10^{-3}$\\
Epochs / batch size & 30 / 1024\\
Discount $\gamma$ (default / post hoc) & 0.95 / 0.99\\
Target update $\tau$ & 0.005\\
\bottomrule
\end{tabular}
\end{table}

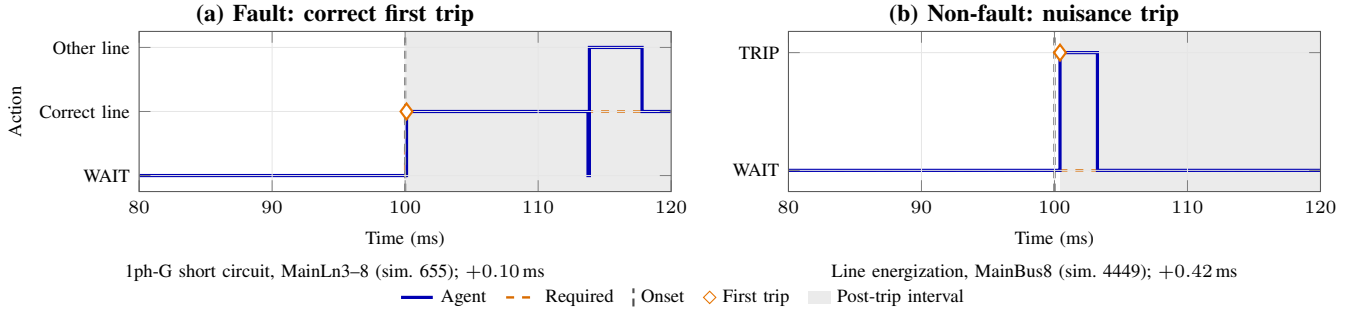
\begin{figure*}[t]
\centering

\begin{minipage}[t]{0.49\textwidth}
\centering
{\small\bfseries (a) Fault: correct first trip\par}
\vspace{0.4ex}
\begin{tikzpicture}
\begin{axis}[
  width=0.97\linewidth,
  height=3.7cm,
  xmin=80, xmax=120,
  ymin=-0.25, ymax=2.25,
  xtick={80,90,100,110,120},
  ytick={0,1,2},
  yticklabels={WAIT,Correct line,Other line},
  xlabel={Time (ms)},
  ylabel={Action},
  tick label style={font=\scriptsize},
  label style={font=\scriptsize},
  xmajorgrids,
  ymajorgrids,
  grid style={black!10,line width=0.2pt},
  axis line style={black!65},
  axis on top,
]

\fill[black!8]
  (axis cs:100.1042,-0.25)
  rectangle
  (axis cs:120,2.25);

\addplot[
  orange!85!black,
  dashed,
  line width=0.9pt
]
coordinates {
  (80.0000,0)
  (100.0000,0)
  (100.0000,1)
  (120.0000,1)
};

\addplot[
  blue!70!black,
  line width=1.2pt
]
coordinates {
  (80.0000,0)
  (100.1042,0)
  (100.1042,1)
  (113.7500,1)
  (113.7500,0)
  (113.8542,0)
  (113.8542,2)
  (117.8125,2)
  (117.8125,1)
  (120.0000,1)
};

\draw[
  black!70,
  densely dashed,
  line width=0.8pt
]
  (axis cs:100.0000,-0.25)
  --
  (axis cs:100.0000,2.25);

\addplot[
  orange!90!black,
  only marks,
  mark=diamond*,
  mark options={
    fill=white,
    line width=0.8pt
  },
  mark size=3pt
]
coordinates {
  (100.1042,1)
};

\end{axis}
\end{tikzpicture}
\vspace{-0.5ex}
{\scriptsize
1ph-G short circuit, MainLn3--8 (sim.~655); $+0.10$\,ms\par}
\end{minipage}
\hfill
\begin{minipage}[t]{0.49\textwidth}
\centering
{\small\bfseries (b) Non-fault: nuisance trip\par}
\vspace{0.4ex}
\begin{tikzpicture}
\begin{axis}[
  width=0.97\linewidth,
  height=3.7cm,
  xmin=80, xmax=120,
  ymin=-0.18, ymax=1.18,
  xtick={80,90,100,110,120},
  ytick={0,1},
  yticklabels={WAIT,TRIP},
  xlabel={Time (ms)},
  tick label style={font=\scriptsize},
  label style={font=\scriptsize},
  xmajorgrids,
  ymajorgrids,
  grid style={black!10,line width=0.2pt},
  axis line style={black!65},
  axis on top,
]

\fill[black!8]
  (axis cs:100.4167,-0.18)
  rectangle
  (axis cs:120,1.18);

\addplot[
  orange!85!black,
  dashed,
  line width=0.9pt
]
coordinates {
  (80.0000,0)
  (120.0000,0)
};

\addplot[
  blue!70!black,
  line width=1.2pt
]
coordinates {
  (80.0000,0)
  (100.4167,0)
  (100.4167,1)
  (103.2292,1)
  (103.2292,0)
  (120.0000,0)
};

\draw[
  black!70,
  densely dashed,
  line width=0.8pt
]
  (axis cs:100.0000,-0.18)
  --
  (axis cs:100.0000,1.18);

\addplot[
  orange!90!black,
  only marks,
  mark=diamond*,
  mark options={
    fill=white,
    line width=0.8pt
  },
  mark size=3pt
]
coordinates {
  (100.4167,1)
};

\end{axis}
\end{tikzpicture}
\vspace{-0.5ex}
{\scriptsize
Line energization, MainBus8 (sim.~4449); $+0.42$\,ms\par}
\end{minipage}

\par\vspace{1.0ex}

{\scriptsize
\tikz[baseline=-0.5ex]
  \draw[blue!70!black,line width=1.2pt](0,0)--(0.42,0);
Agent\quad
\tikz[baseline=-0.5ex]
  \draw[orange!85!black,dashed,line width=0.9pt](0,0)--(0.42,0);
Required\quad
\tikz[baseline=-0.5ex]
  \draw[black!70,densely dashed,line width=0.8pt](0,-0.12)--(0,0.12);
Onset\quad
\tikz[baseline=-0.5ex]
  \node[diamond,draw=orange!90!black,fill=white,
        inner sep=1.2pt] {};
First trip\quad
\tikz[baseline=-0.5ex]
  \fill[black!8](0,-0.08)rectangle(0.30,0.08);
Post-trip interval
}

\caption{Representative held-out action sequences for the default
combined-input $W=48$ policy. (a) A correct first trip fixes the terminal
fault outcome despite a later wrong-line prediction. (b) A single nuisance
trip fixes a false non-fault outcome despite predominantly correct waiting.
Shading marks predictions ignored by the first-trip evaluation but retained
in the dense metrics.}
\label{fig:terminal-examples}
\end{figure*}

\section{Experimental Protocol}
\label{sec:experimental_protocol}

The controlled sensitivity study has three blocks. Block A compares phasor and combined representations at $W=48$ and 96 with $\alpha=0.5$. Block B uses combined input and $W=48$: two variants change the false-positive penalty to $-10$ and $-200$, while a third retains the default penalties and increases the shared correct reward from 5 to 50 for both correct fault trips and post-event non-fault waits. Block C changes $\alpha$ to 0.1 and 0.9 around the default 0.5. Each run uses the same partitions, inputs, seed, 30-epoch stopping point, and evaluation episodes. Following review of the initial results, we conducted one exploratory post-hoc run using combined input with $W=48$ and $\gamma=0.99$. It otherwise retains the inputs, split, seed, architecture, reward, CQL weight, and 30-epoch stopping point. The run is reported separately from the prespecified three-block matrix.

Every run saves its configuration, environment, training history, checkpoints, predictions, metrics, and file hashes. Checkpoints retain Python, NumPy, and CPU random-number-generator state so that interrupted seeded training resumes consistently. Evaluation uses the final epoch for every configuration; neither dense scores nor first-trip outcomes guide early stopping or checkpoint selection.

\subsection{Per-Timestep Evaluation}

After the fault onset, the correct line-trip action is a \ac{tp}; waiting or selecting another line-trip action is a \ac{fn}. Before fault onset and throughout non-fault episodes, waiting is a \ac{tn} and any trip is a \ac{fp}. We report precision, recall, F1-score, and \ac{fpr}. Recall quantifies how reliably the policy selects the faulted line after fault onset, whereas \ac{fpr} quantifies unnecessary trips before event onset or during non-fault episodes. These metrics characterize dense action predictions, not a deployed terminal relay. To exclude incomplete phasor windows, evaluation begins only once the full causal support is available: after 239 raw samples for $W=48$ and 287 raw samples for $W=96$. The post-fault interval is identical, but the longer-window configurations contain 48 fewer pre-event decisions; their \ac{fpr} and first-trip security therefore use a slightly shorter negative-event horizon.

\subsection{Terminal First-Trip Evaluation}

For each saved sequence of evaluation predictions, the first action other than \emph{wait} is the first trip; all later actions are ignored. In a fault episode, it is classified as a premature trip, correct-line trip, wrong-line trip, or right-censored no-trip. In a non-fault episode, it is classified as a false trip or no-trip. For correct post-fault trips, latency is
\begin{equation}
t_{\mathrm{lat}}=\frac{n_{\mathrm{trip}}-960}{9600}\,\mathrm{s}.
\label{eq:latency}
\end{equation}
Wilson 95\,\% confidence intervals are reported for the highlighted non-fault false-trip rates. Correct-trip latency is summarized by its median and 95th percentile. The terminal evaluation definition was fixed before computing the results.

\section{Results}
\label{sec:results}

Table~\ref{tab:step} reports the nine prespecified runs and the exploratory post-hoc discount-factor run. Window length does not have a representation-independent effect. Phasor-only F1 decreases from 0.9632 at $W=48$ to 0.9454 at $W=96$, whereas combined-input F1 increases from 0.9637 to 0.9686. Combining raw and phasor data is therefore nearly neutral at $W=48$ but improves F1 by 0.0231 at $W=96$. The longer combined model also reduces \ac{fpr} from 2.05\,\% to 0.32\,\%.

\begin{table*}[ht]
\centering
\caption{Per-timestep evaluation performance. P, R, and F1 are proportions; FPR is reported in percent.}
\label{tab:step}
\begin{tabular}{llrrrrrr}
\toprule
\textbf{Block} & \textbf{Configuration} & $\mathbf{W}$ & $\boldsymbol{\alpha}$ & \textbf{P} & \textbf{R} & \textbf{F1} & \textbf{FPR (\%)}\\
\midrule
A & Phasor & 48 & 0.5 & 0.9922 & 0.9359 & 0.9632 & 2.97\\
A & Combined & 48 & 0.5 & 0.9946 & 0.9347 & 0.9637 & 2.05\\
A & Phasor & 96 & 0.5 & 0.9992 & 0.8972 & 0.9454 & 0.31\\
A & Combined & 96 & 0.5 & 0.9992 & 0.9397 & 0.9686 & 0.32\\
\midrule
B & Combined, FP penalty $-10$ & 48 & 0.5 & 0.9946 & 0.9347 & 0.9637 & 2.05\\
B & Combined, FP penalty $-200$ & 48 & 0.5 & 0.9946 & 0.9347 & 0.9637 & 2.05\\
B & Combined, correct reward 50 & 48 & 0.5 & 0.9996 & 0.9252 & 0.9610 & 0.13\\
\midrule
C & Combined & 48 & 0.1 & 0.9977 & 0.8236 & 0.9023 & 0.76\\
C & Combined & 48 & 0.9 & 0.9993 & 0.9496 & \textbf{0.9738} & 0.28\\
Post hoc & Combined, $\gamma=0.99$ & 48 & 0.5 & 0.9997 & 0.9422 & 0.9701 & 0.10\\
\bottomrule
\end{tabular}
\end{table*}

Changing only the false-positive penalty has no measured effect: the $-10$, default $-100$, and $-200$ settings yield identical predictions and metrics. The sampled transition archive contains no pre-event or non-fault trip rows, so this parameter changes none of the sampled training rewards. Increasing the shared correct reward from 5 to 50 affects both correct fault trips and post-event non-fault waits; it lowers \ac{fpr} to 0.13\,\% but also lowers recall and F1. No monotonic reward conclusion is supported.

Performance is sensitive to the tested \ac{cql} weight. Reducing $\alpha$ from 0.5 to 0.1 lowers recall to 0.8236 and F1 to 0.9023. Increasing it to 0.9 gives the highest recall (0.9496) and F1 (0.9738), with an \ac{fpr} of 0.28\,\%. Because the study does not include $\alpha=0$, these results characterize sensitivity only within the tested \ac{cql} range.
Compared with the default combined-$W=48$ model, the exploratory $\gamma=0.99$ run improves precision from 0.9946 to 0.9997, recall from 0.9347 to 0.9422, and F1-score from 0.9637 to 0.9701, while reducing FPR from 2.05\,\% to 0.10\,\%. It remains below the $\alpha=0.9$ model's F1-score; one post-hoc setting cannot establish a general trend.

\subsection*{First-Trip Behavior}

The terminal results in Table~\ref{tab:first-trip} differ from the per-timestep ranking. Figure~\ref{fig:terminal-examples} illustrates the asymmetry of this evaluation: later prediction errors cannot undo an earlier correct trip, whereas a single nuisance trip determines the terminal outcome of a non-fault episode. All configurations have zero premature trips. The default combined $W=48$ agent selects the correct line-trip action first in 210 of 214 fault episodes (98.13\,\%), with one wrong-line trip and three no-trip episodes. Its median correct-trip latency is one timestep (0.104~ms), and its 95th percentile is 1.667~ms. The post-hoc $\gamma=0.99$ model is correct first in 211 fault episodes (98.60\,\%), with one wrong-line trip and two no-trip episodes; its median latency remains 0.104~ms, while its 95th percentile increases to 1.927~ms. By contrast, the per-timestep F1 leader at $\alpha=0.9$ is correct first in 197 fault episodes (92.06\,\%), selects a wrong-line trip in 15, and has two no-trip episodes. This distinction follows directly from terminal semantics: a later wrong action cannot undo a correct first trip, while an early wrong-line trip cannot be repaired by subsequent predictions.

\begin{table}[ht]
\centering
\caption{Terminal first-trip evaluation performance. Fault outcomes are percentages of 214 episodes. NF trip is the false-trip percentage among 11 non-fault episodes. $p_{95}$ is correct-trip latency in ms.}
\label{tab:first-trip}
\setlength{\tabcolsep}{2.8pt}
\begin{tabular}{lrrrrr}
\toprule
\textbf{Configuration} & \textbf{Correct} & \textbf{Wrong line} & \textbf{No trip} & \textbf{NF trip} & \textbf{$p_{95}$}\\
\midrule
Phasor $W=48$ & 87.38 & 12.15 & 0.47 & 72.73 & 1.219\\
Combined $W=48$ & 98.13 & 0.47 & 1.40 & 72.73 & 1.667\\
$\gamma=0.99$ (post hoc) & \textbf{98.60} & 0.47 & 0.93 & 54.55 & 1.927\\
Phasor $W=96$ & 94.86 & 5.14 & 0.00 & 45.45 & 1.354\\
Combined $W=96$ & 89.72 & 9.81 & 0.47 & 72.73 & 1.042\\
FP penalty $-10$ & 98.13 & 0.47 & 1.40 & 72.73 & 1.667\\
FP penalty $-200$ & 98.13 & 0.47 & 1.40 & 72.73 & 1.667\\
Correct reward 50 & 97.66 & 0.93 & 1.40 & 54.55 & 2.125\\
$\alpha=0.1$ & 93.46 & 0.47 & 6.07 & \textbf{36.36} & 6.896\\
$\alpha=0.9$ & 92.06 & 7.01 & 0.93 & 54.55 & 2.750\\
\bottomrule
\end{tabular}
\end{table}

Non-fault security remains the main limitation. The default combined $W=48$ agent trips in 8 of 11 non-fault episodes (72.73\,\%; Wilson 95\,\% CI 43.44--90.25\,\%), despite its per-timestep FPR of 2.05\,\%. The lowest observed nuisance-trip rate occurs for $\alpha=0.1$, with 4 of 11 episodes (36.36\,\%; CI 15.17--64.62\,\%). This apparent security improvement is accompanied by only 200 correct first trips among 214 fault episodes, 13 no-trip fault episodes, and the largest $p_{95}$ latency of 6.896~ms; it is therefore consistent with greater reluctance to trip rather than uniformly better protection. The $\alpha=0.9$ and post-hoc $\gamma=0.99$ models each trip in 6 of 11 non-fault episodes (54.55\,\%; CI 28.01--78.73\,\%). In the paired discount-factor comparison, $\gamma=0.99$ removes the nuisance trip in two episodes and introduces none, while correct first trips on fault episodes increase from 210 to 211. Nevertheless, the 11 non-fault episodes, one training seed, and post-hoc design preclude a general conclusion that a larger $\gamma$ improves security.

\subsection*{Fault-Family Analysis}

Table~\ref{tab:fault-family} groups the first-trip outcomes into conventional short circuits, \ac{hif}, and incipient faults. The default combined $W=48$ agent is correct first on all 193 conventional short-circuit episodes, but only on 11 of 13 \ac{hif} and 6 of 8 incipient-fault episodes. The post-hoc $\gamma=0.99$ model is correct first in 192 of 193 conventional short-circuit episodes, all 13 \ac{hif} episodes, and 6 of 8 incipient-fault episodes. Its net gain of one correct fault episode therefore comprises two additional correct \ac{hif} operations offset by one fewer correct conventional short-circuit operation. The small \ac{hif} and incipient-fault counts preclude strong conclusions. They nevertheless contradict a blanket conclusion that incipient faults are always missed: several policies select the correct line-trip action first in most or all of these eight episodes.

\begin{table}[ht]
\centering
\caption{Correct-first-trip rate (\%) by fault family. Parentheses give the episode count shared by every row.}
\label{tab:fault-family}

\setlength{\tabcolsep}{3.4pt}
\begin{tabular}{lrrr}
\toprule
\textbf{Configuration} & \textbf{Short circuit (193)} & \textbf{\ac{hif} (13)} & \textbf{Incipient (8)}\\
\midrule
Phasor $W=48$ & 89.64 & 61.54 & 75.00\\
Combined $W=48$ & \textbf{100.00} & 84.62 & 75.00\\
Phasor $W=96$ & 95.34 & 84.62 & \textbf{100.00}\\
Combined $W=96$ & 93.78 & 38.46 & 75.00\\
Combined $\alpha=0.9$ & 93.26 & 92.31 & 62.50\\
$\gamma=0.99$ (post hoc) & 99.48 & \textbf{100.00} & 75.00\\
\bottomrule
\end{tabular}
\end{table}

\section{Discussion and Limitations}
\label{sec:discussion}

Per-timestep and terminal metrics support different conclusions. Phasor features were sufficient for strong performance, while raw waveforms helped only with the longer observation window. The tested \ac{cql} weight affected performance, yet the best dense F1 did not produce the best terminal behavior. Reward conclusions were also limited by actions absent from the fixed archive. This gap between nominal and operational rankings complements prior findings that clean-data performance may not predict behavior under degraded measurements~\cite{oelhaf_robustness_2026}.

The post-hoc discount-factor result has a plausible but limited explanation. Wait actions receive bootstrapped future value, whereas trip actions terminate a sampled transition, so increasing $\gamma$ can favor waiting. However, the recorded next state is action-independent and the archive contains no pre-event or non-fault trip rows. The nuisance-trip reduction is therefore an indirect policy effect, not evidence that a larger discount factor generally solves the security problem.

These experiments establish feasibility, not deployment readiness. The static trajectories do not respond to trip actions, and development and evaluation cover one simulated network and operating distribution. Only 11 non-fault episodes are available, first-trip evaluation omits breaker dynamics, coordination time, communication delay, hardware uncertainty, and distribution shifts, and each configuration uses one training seed. The reported intervals therefore do not quantify seed variability or the low false-trip probabilities required in practice.

Future work should test frozen policies in interactive or controller-hardware-in-the-loop settings with terminal breaker actions, broader non-fault coverage, topology and contingency shifts, and explicit security targets. Conventional-relay baselines and calibrated supervisory logic are needed. Dense metrics should be retained for diagnostic tracking, but model selection should optimize terminal protection costs because the first trip represents the irreversible relay command.

\section{Conclusion}
\label{sec:conclusion}

We studied offline \ac{cql} for line-selective tripping on simulated CIGRE medium-voltage trajectories using causal phasor and impedance features. Across the controlled sensitivity study, combined input with $W=96$ achieved an F1-score of 0.9686, while increasing the \ac{cql} weight to 0.9 at $W=48$ gave the best per-timestep F1-score of 0.9738. Window and representation effects were coupled, while the false-positive-penalty variants were identical because the sampled archive contains no transitions on which that penalty is applied.

Terminal evaluation changed the interpretation: the default combined $W=48$ model selected the correct line-trip action first in 98.13\,\% of fault episodes, yet false-tripped in 72.73\,\% of the small non-fault set. The exploratory $\gamma=0.99$ follow-up reaches an F1-score of 0.9701 and the lowest dense FPR, 0.10\,\%, while changing the terminal outcomes from 210 to 211 correct fault trips and from eight to six non-fault trips. These changes are encouraging, but the small non-fault set and single-seed post-hoc design do not support a general discount-factor conclusion. Overall, offline \ac{cql} yields strong faulted-line selection on these simulations, but substantially broader security testing and an interactive environment are required before the approach can be considered for practical grid protection.

\vfill

\section*{Data and Code Availability}
The EvEMTBench data used in this study are publicly available via
\href{https://data.fau.de/share/0e8d60feb7e65616c60aab78b93db77053275da53fd894bf5b75fc5e9ee7dfbf/}
{FAUDataCloud}.
Code for data preparation, training, evaluation, and reproduction is available at
\href{https://github.com/julianoelhaf/offline-cql-protection}
{github.com/julianoelhaf/offline-cql-protection}.

\section*{Acknowledgment}
This project was funded by the Deutsche Forschungsgemeinschaft (DFG, German Research Foundation) - 535389056.

\bibliographystyle{IEEEtran}
\bibliography{references}

\vspace{0.5cm}

\end{document}